\documentclass{nature}

\title{Ozone observations reveal lower solar cycle spectral variations}

\author{Ball, W.T.$^{1}$, Haigh, J.D.$^{2}$, Rozanov, E.V.$^{1,3}$, Kuchar, A.$^{3,4}$, Sukhodolov, T.$^{1,3}$, Tummon, F.$^{3}$, Shapiro, A.V.$^{1}$ \& Schmutz, W.$^{1}$}
 
\begin{document}

\maketitle

\begin{affiliations}
 \item Physikalisch-Meteorologisches Observatorium Davos World Radiation Centre, Dorfstrasse 33, 7260 Davos Dorf, Switzerland
 \item Grantham Institute - Climate Change and the Environment, Imperial College London, South Kensington Campus, London SW7 2AZ, UK
 \item Institute for Atmospheric and Climate Science, Swiss Federal Institute of Technology Zurich, Universitaetstrasse 16, CHN, CH-8092 Zurich, Switzerland
 \item Department of Atmospheric Physics, Faculty of Mathematics and Physics, Charles University in Prague, V Holesovickach 2, 180 00 Prague 8, Czech Republic
\end{affiliations}

\textbf{Some of the natural variability in climate is understood to come from changes in the Sun. A key route whereby the Sun may influence surface climate is initiated in the tropical stratosphere by the absorption of solar ultraviolet (UV) radiation by ozone, leading to a modification of the temperature and wind structures and consequently to the surface through changes in wave propagation and circulation~\cite{Haigh1996,KoderaKuroda2002,SimpsonBlackburn2009,GrayBeer2010}. While changes in total, spectrally-integrated, solar irradiance lead to small variations in global mean surface temperature, the `top-down' UV effect preferentially influences on regional scales at mid-to-high latitudes with, in particular, a solar signal noted in the North Atlantic Oscillation~\cite{ShindellSchmidt2001,GrayScaife2013} (NAO). The amplitude of the UV variability is fundamental in determining the magnitude of the climate response but understanding of the UV variations has been challenged recently by measurements from the SOlar Radiation and Climate Experiment (SORCE) satellite~\cite{Rottman2005}, which show UV solar cycle changes up to 10 times larger than previously thought. Indeed, climate models using these larger UV variations show a much greater response, similar to NAO observations~\cite{InesonScaife2011}. Here we present estimates of the ozone solar cycle response using a chemistry-climate model (CCM) in which the effects of transport are constrained by observations. Thus the photolytic response to different spectral solar irradiance (SSI) datasets can be isolated. Comparison of the results with the solar signal in ozone extracted from observational datasets yields significantly discriminable responses. According to our evaluation the SORCE UV dataset is not consistent with the observed ozone response whereas the smaller variations suggested by earlier satellite datasets, and by UV data from empirical solar models, are in closer agreement with the measured stratospheric variations. Determining the most appropriate SSI variability to apply in models will allow clearer understanding of the impact of past and future solar variability on global and regional climate.}


Solar UV radiation at wavelengths shorter than 242 nm initiate the creation of ozone in the middle atmosphere while wavelengths shorter than 320 nm destroy it. It follows that the distribution of ozone is sensitive to spectral solar irradiance (SSI) and, conversely, that observed changes in ozone can reveal information on variations in SSI \cite{ErmolliMatthes2013,BallKrivova2014,BallMortlock2014} (see Fig. 1). This is especially true in the tropical upper stratosphere where photochemical processes dominate over transport. 

Measurements of the solar UV spectrum have been acquired since 2003 by two instruments on the SORCE satellite: the SOLar Stellar Irradiance Comparison Experiment (SOLSTICE)~\cite{McClintockRottman2005} and the Spectral Irradiance Monitor (SIM)~\cite{HarderFontenla2009}. The very large solar cycle variations shown in early versions of the data, probably due to insufficient correction of sensor degradation and drifts~\cite{ErmolliMatthes2013}, have reduced in recent releases, but significant uncertainties remain over the true magnitude of UV variation (see, e.g., Fig.~3 of ref [10] and Fig.~4 of ref [11]).

The total solar irradiance (TSI) amplitude of the current solar cycle is $\sim$65\% of the previous cycle, and TSI~\cite{YeoKrivova2014} and open solar flux~\cite{Lockwood2013} have decreased between recent solar minima following a century in a grand maximum state~\cite{UsoskinSolanki2012}. It has been suggested~\cite{FontenlaCurdt2009} that the larger SORCE UV trends may indicate that the Sun's spectral variability may have been behaving differently during the last solar cycle. TSI and 27-day solar rotational variability is well constrained by observations and models reproduce these well. Sunspots and faculae drive irradiance variations on daily to centennial timescales. So, for the solar cycle spectral trends to change between solar cycles, the wavelength dependent intensities of the sunspots and faculae would also have to change~\cite{HarderFontenla2009} in such a way that TSI on all timescales and SSI on rotational timescales does not change. This is highly unlikely, but does not preclude the possibility that the SORCE magnitude is correct.

To investigate SORCE trends over more than half a solar cycle, we extrapolate SORCE UV solar cycle trends back to 1974 using the SATIRE-S model while ensuring agreement with TSI (see Methods). SATIRE-S is a semi-empirical solar model~\cite{YeoKrivova2014} constructed using time-independent model intensities of sunspots, faculae and the quiet-Sun. It agrees with TSI observations better than any other model, reproduces rotational variability well and shows good agreement with UARS Solar Ultraviolet Spectral Irradiance Monitor (SUSIM; 1991-2005) solar cycle trends below 400 nm~\cite{BallKrivova2014}. The empirical Naval Research Laboratory spectral model~\cite{LeanRottman2005} ('NRLSSI'), typically used in climate studies, displays slightly lower solar cycle trends than SATIRE-S above 250 nm~\cite{BallKrivova2014} (Fig.~2a). Figure 2a shows SATIRE-S, NRLSSI and extrapolated SORCE ('eSORCE') from 1991 to 2012 integrated over 250-300 nm. Interestingly, from 2009 onwards SORCE/SIM UV agrees well with SATIRE-S.

The solar cycle ozone response is usually extracted from ozone observations using multi-linear regression (MLR; see Methods). The results have, however, varied widely. Early analyses generally showed a positive correlation throughout the middle atmosphere, with a peak solar cycle amplitude of ~2\% at 40 km ($\sim$5 hPa)~\cite{AustinTourpali2008}.  More recently some studies have suggested a negative relationship in the lower mesosphere~\cite{HaighWinning2010,MerkelHarder2011,BallMortlock2014} while others have not~\cite{HoodMisios2015}. Results from experiments with atmospheric models have shown that the negative signal is consistent with the larger UV variations from earlier versions of SORCE data, although the predicted magnitude reduces with updated versions~\cite{BallKrivova2014}.

Uncertainties associated with both the MLR analyses (e.g. the possibility of aliasing between input proxies~\cite{ChiodoMarsh2014}) and the model simulations (e.g. their representation of transport processes and how these respond to the Sun) mean that it has not been possible to narrow the uncertainty range of SSI variations. To circumvent these issues we calculate ozone changes between 1991 and 2012 using the full photochemical capabilities of the SOlar Climate Ozone Links (SOCOL) chemistry-climate model (CCM)~\cite{StenkeSchraner2013} together with nudged dynamical fields from ERA-Interim reanalysis data (see Methods). We perform three simulations, the first using a solar-minimum constant-Sun. Zonal-mean tropical ozone at 1.6 hPa from that simulation are shown by the red curve in Fig. 2b and compared with the Stratospheric Water and OzOne Satellite Homogenized (SWOOSH) ozone observations dataset~\cite{DavisEA2015} in black.  The model reproduces the overall magnitude of the semi-annual variation and inter-annual variability shown by the measurements very well. The difference between the two curves, shown in Fig. 2c by the grey curve (with a 24-month running mean in black and 1$\sigma$ uncertainty range by shading), clearly indicates a solar cycle signal in the observed data. 

The other two model simulations use varying SSI from the SATIRE-S model and from eSORCE, as exemplified in Fig. 2a. The differences between these simulations and the constant-Sun run are shown in Fig. 2c by the blue and green curves respectively.  Both show a solar cycle influence although SATIRE-S is of smaller magnitude than SWOOSH and eSORCE is inverted at this altitude, consistent with the larger UV variations.

We perform MLR (see Methods) throughout the 30-0.25 hPa tropical region; the nudged simulations provide a clean output eliminating the need for ensemble simulations to detect the solar signal. We perform the same analysis on the SWOOSH and the Global OZone Chemistry And Related Datasets for the Stratosphere (GOZCARDS)~\cite{FroidevauxAnderson2015} ozone datasets. Fig.~3 shows that the simulation using SATIRE-S is more like the results from the measured data than that using eSORCE. 


To further test the robustness of these results we compare the photolytic trends over two solar maximum-to-minimum periods: 1991/07 - 1996/05 and 2002/02 - 2008/12. We remove the dynamical response by subtracting the constant-Sun simulation from the other simulations and observed merged ozone datasets (as in Fig.~2c), leaving a residual photolytic response. Additionally, we include data from SBUV-Merged and SBUV-Mod~\cite{McPetersBhartia2013} ozone datasets in the latter period, but not prior to 2002 due to steps in the data where there is a change in the underlying instrument data used; SWOOSH has too many data gaps to consider the 1996 - 2002 period (see Fig.~2c). We calculate the trends and uncertainties from the residuals (see methods). The results are shown in Fig.~4. Caution is needed for results above 1 hPa; the large diurnal cycle in ozone (up to 10\%~\cite{FroidevauxAnderson2015}) means that observations from different times of the day need to be adjusted to ensure this is taken into account: SWOOSH does not extend beyond 1 hPa, the GOZCARDS authors advise caution~\cite{FroidevauxAnderson2015}; and the SBUV merged datasets do not correct for it~\cite{TummonHassler2014}.

All four ozone datasets show broadly the same solar cycle ozone response, peaking at $\sim$4 hPa; the SBUV responses are slightly smaller than GOZCARDS and SWOOSH above 10 hPa and are not significantly different from zero above 2 hPa. GOZCARDS and SWOOSH agree with each other in both time periods, except at 4 hPa where the earlier period has a larger response. The fact that the solar cycle ozone responses are similar in both solar cycles indicates that the solar cycle UV changes were also similar in both periods. This suggests that it is unlikely that the behaviour of the Sun's spectral variability changed between the cycles.

Fig.~4 also shows SORCE and SATIRE-S simulation results as green and blue shading, respectively (shading encompasses the trend fitting and errors, which are similar in both periods). The actual solar cycle response, calculated by taking the difference of six-month means at solar maximum and solar minimum, agree within the uncertainty of the trend fitting, showing that the linear fitting is appropriate. The red shading represents NRLSSI for 2002-2008 only. The merged ozone datasets, for both periods, reveal statistically significant results that do not agree with the simulation using SORCE above $\sim$4hPa and, rather, have similar ozone profiles to those using SATIRE-S and NRLSSI.

The simulation profile shapes in Fig.~4 differ from Fig.~3 because the dynamical solar response is included; the lower altitude positive response and negative response higher up are indicative of a slowing of the Brewer-Dobson circulation in agreement with theory~\cite{KoderaKuroda2002}. Subtracting the constant-Sun (not shown) from the SATIRE-S and SORCE profiles give photolytic responses similar in shape and magnitude to those in Fig.~4.

Overall, we find that the ozone observations are not consistent with SORCE UV solar cycle changes. Our novel analysis provides a unique assessment of spectral solar cycle changes and strongly supports suggestions~\cite{ErmolliMatthes2013} that SORCE solar cycle trends at longer UV wavelengths, especially above 250 nm, are unlikely to be correct. Our study also shows that SATIRE-S spectral changes, which are similar to older observations from UARS/SUSIM, produce ozone changes more consistent with observations.

Solar UV variability influences the climate near Earth's surface through heating the middle atmosphere and subsequent dynamical coupling between the stratosphere and troposphere. Details of the changes in the UV spectrum are crucial to the mechanisms involved and to the resulting impacts. A recent study~\cite{InesonMaycock2015}, concerned with the potential for declining solar activity to mitigate global warming, has shown that the effect on near-surface temperature in the North Atlantic depends sensitively on the choice of UV spectrum and our work suggests that the effect is probably at the lower end of those considered in that study. If climate models are not able to reproduce the observed signals at the surface using the weaker UV changes it may be that they are missing some necessary mechanism(s). A better knowledge of SSI variability is essential to advance our understanding of how the Sun influences surface climate and might do so in future.

\begin{figure}
\caption{\textbf{Illustration of photolytic solar cycle ozone response.} The solar maximum to minimum photolytic ozone response from integrated-UV below 242 nm (left column) is to produce ozone. Wavelengths below 320 nm (middle) photolyse O$_{3}$ leading to a catalytic loss of O$_{3}$ if the oxygen atom does not recombine with O$_{2}$. These responses add almost linearly to give the resultant solar cycle response~\cite{BallMortlock2014} (right). The far larger SORCE changes at ozone destroying wavelengths more than compensates for the larger UV changes at shorter wavelengths, leading to a negative response higher up, while SATIRE-S maintains a positive, but smaller, response at all altitudes.}
\end{figure}

\begin{figure}
\caption{\textbf{Solar irradiance and ozone timeseries.} (\textbf{a}) Daily and smoothed spectral irradiance integrated over 250-300 nm for SATIRE-S (blue) and NRLSSI (red, smoothed only) models, and SORCE/SIM (black) and extrapolated SORCE (green; extrapolated from the period between the dashed lines) observations; the bar represents 1\% variability. (\textbf{b}) 1.6 hPa ($\sim$43 km) zonal ozone (20$^{\circ}$S-20$^{\circ}$N averaged) from SWOOSH (black) and constant-Sun simulation. (\textbf{c}) Residuals from the constant-Sun simulation for SWOOSH (monthly, grey; 24-month running mean, black; 1$\sigma$ uncertainty, shading), SATIRE-S (blue) and extrapolated-SORCE (green). Data are bias-shifted to the constant-simulation solar minimum around 2008. Vertical bars indicate solar maxima (solid) and minima (dotted).}
\end{figure}

\begin{figure}
\caption{\textbf{Ozone response from multi-linear regression between 1991 and 2012.} Multi-linear regression solar cycle mean responses and 2$\sigma$ uncertainties for SWOOSH and GOZCARDS ozone composites (dot-bar) and for SORCE (green shading) and SATIRE-S (blue shading) simulations. The ozone response is in term of 100 solar flux units of the 10.7 cm radio flux, about 80\% of the SC.}
\end{figure}

\begin{figure}
\caption{\textbf{Solar cycle maximum to minimum ozone response from linear fitting.} The extracted solar cycle change in ozone using Theil-Sen trend analysis (2$\sigma$ error bars) from the residuals of the constant-Sun simulation (e.g. in Fig 2c) with ozone time series from 1991/07 to 1996/05 (dashed, circles) and 2002/02 to 2008/12 (solid, filled circles). SWOOSH (black) and GOZCARDS (orange) are given for both periods; SBUV-Merged (pink) and SBUV-Mod (purple) are given for the latter period. SATIRE-S (blue) and SORCE (green) shading combines the uncertainty ranges from both periods with the change between six-month averages at maximum and minimum; NRLSSI is also shown for the latter period.} 
\end{figure}

\begin{methods}

\subsection{SSI datasets.}: Four SSI datasets were used: constant-Sun using the mean of 2008/11 - 2009/01 from SATIRE-S; SATIRE-S model; NRLSSI model; extrapolated-SORCE. Each dataset was bias corrected to give the same absolute fluxes as SATIRE-S at the solar minimum in 2008.

\subsection{Extrapolated-SORCE (eSORCE) solar irradiance data.} SORCE are extrapolated using SATIRE-S to the 1974-2013 period. Smoothed SORCE data are regressed to SATIRE-S between 2004/09/01 and 2008/08/31 (dashed vertical lines in Fig.~2) and SATIRE-S solar cycle trends were then scaled to SORCE. Prior to 2004/09 SORCE/SIM UV shows a change in the cycle gradient so it is not considered. We use UV SORCE data from SOLSTICE v13 below 247 nm and SIM v20 between 247 and 310 nm; changeover at 247 nm was chosen due to higher correlations of SORCE/SIM with SATIRE-S here. SATIRE-S fluxes are used above 1598 nm. Integrated SORCE data are not consistent with total solar irradiance (TSI) so, to conserve integrated-eSORCE with TSI observations, the 310-1598 nm fluxes were rescaled by a factor of 0.15. SATIRE-S rotational variability agrees well with observations and is added to the solar cycle trend of eSORCE to give daily variability. We note that SIM v22 and SOLSTICE v14 show nearly identical solar cycle trends for the 2004-2009 extrapolation period.

\subsection{Ozone Data}
A summary of the ozone merged datasets, and an intercomparison, are given by [28]. These data are monthly, zonally averaged, homogenised, and bias-corrected ozone datasets spanning 1984-2013, typically covering latitudes between 48 S and 48 N. All datasets were interpolated onto the SOCOL model pressure levels. Data were then bias corrected to the constant-Sun simulation mean values for the six-month average centred on the December 2008 solar minimum.


\subsection{Nudged Chemistry Climate Model.}
SOCOL uses the ECHAM5 atmospheric model with a resolved stratosphere and online coupled chemistry module (MEZON)~\cite{StenkeSchraner2013}. We use the Stratospheric Processes and their Role in Climate (SPARC) International Global Atmospheric Chemistry (IGAC) Chemistry Climate Model Intercomparison (CCMI) boundary conditions and external forcings (except for the Sun). We nudge the wind, temperature, and surface level pressure with ERA-Interim up to 0.1 hPa between 1983 to 2012, only considering data from 1991 to 2012; 1983 - 1990 are considered as model spin-up years. A run was also performed using the NRLSSI model for 2002-2010; spin-up was for 1994-2001.

\subsection{Trend analysis.}
We perform linear trend analysis using the non-parametric Theil-Sen trend estimation, which is more accurate than a standard linear regression for non-gaussian data.

\subsection{Multi-linear regression.}
We have applied Multiple Linear Regression (MLR) similar to~\cite{KucharSacha2014} using the 10.7 cm radio flux as a solar proxy, stratospheric aerosol optical depth (SAOD) for volcanic eruptions, an ENSO index representing El Ni\~{n}o Southern Oscillation variability, and two modes of the Quasi-Biennial Oscillation. Statistical significant of the regression coefficients was evaluated with a t-test. We found similar results using a more robust 'bias-corrected and accelerated' (BCA) bootstrap percentile method based on 10000 samples, which does not assume the data distribution \textit{a priori}. 

\end{methods}





\begin{addendum}
 \item[Acknowledgements] This work was funded by the SNSF project 149182. We thank A. Stenke and A. Coulon for their advice, comments and help with the model nudging. We thank the GOZCARDS, SWOOSH and SBUV teams for their ozone products. We acknowledge SATIRE-S data from http://www2.mps.mpg.de/ projects/sun-climate/data.html; NRLSSI data from http://lasp.colorado.edu/lisird/; and SORCE data from http://lasp.colorado.edu/home/sorce/.
 \item[Author Contributions] W.T.B. prepared the solar data, performed the model experiments, carried out the linear analysis and interpretation of results. W.T.B and J.D.H wrote the paper. W.T.B, T.S. and E.V.R. designed and set up the model simulations, F.T. provided the ozone datasets and advice on their use, A.V.S. prepared the CCM model nudging and A.K. performed the MLR analysis. J.D.H., E.V.R. and W.S. provided expert advice and discussion on the results.
 \item[Competing Interests] The authors declare that they have no competing financial interests.
 \item[Correspondence] Correspondence and requests for materials should be addressed to W.T.B. (email: william.ball@ pmodwrc.ch).
\end{addendum}


\end{document}